\documentclass[twocolumn]{aastex631}

\usepackage{amsmath}
\usepackage{booktabs}
\newcommand{\Teff}{T_{\rm eff}}
\newcommand{\logg}{\log g}
\newcommand{\feh}{\mathrm{[M/H]}}
\newcommand{\alphafe}{[\alpha/\mathrm{M}]}
\newcommand{\cm}{[\mathrm{C/M}]}
\newcommand{\nm}{[\mathrm{N/M}]}
\newcommand{\dd}{\mathrm{d}}
\newcommand{\Wone}{W_1}

\shorttitle{Armillary}
\shortauthors{Ting \& Saad}

\begin{document}

\title{Armillary: A Label-Free Coordinate System for Stellar Spectra}

\author[0000-0001-5082-9536]{Yuan-Sen Ting}
\affiliation{Department of Astronomy, The Ohio State University, 140 West 18th Avenue, Columbus, OH 43210, USA}
\affiliation{Center for Cosmology and AstroParticle Physics (CCAPP), The Ohio State University, Columbus, OH 43210, USA}
\affiliation{Max-Planck-Institut f\"ur Astronomie, K\"onigstuhl 17, D-69117 Heidelberg, Germany}

\author[0009-0004-9592-2311]{Serat Saad}
\affiliation{Department of Astronomy, The Ohio State University, 140 West 18th Avenue, Columbus, OH 43210, USA}
\affiliation{Center for Cosmology and AstroParticle Physics (CCAPP), The Ohio State University, Columbus, OH 43210, USA}

\begin{abstract}

A stellar spectrum is determined by a few parameters, so a set of spectra lies on a low-dimensional manifold with those parameters as its coordinates. We introduce Armillary, which generalises the Sequencer of Baron and M\'enard from an ordering to multidimensional coordinates, constructed from the spectra alone. Two stars are compared by the Wasserstein distance between their signed absorption profiles. Shortest paths through a nearest-neighbour graph measure distance along the manifold and give the coordinates, and the same graph transfers the labels of a few stars to all the others. The coordinates are more nearly linear in $\Teff$ and $\logg$ than principal components, t-SNE, UMAP or an autoencoder. The label transfer, without deep learning, is more precise than the first three and comparable to the autoencoder. On $202\,932$ APOGEE DR17 stars, an illustrative draw of $32$ training stars recovers the ASPCAP $\Teff$, $\logg$ and $\feh$ of held-out dwarfs and giants, with $1\sigma$ errors of $100$\,K, $0.18$ and $0.083$\,dex. Resolving the two $\alpha$ sequences of the disc in $\alphafe$ for giants takes $128$ training stars, and with $2048$ the four errors improve to $33$\,K, $0.06$, $0.026$ and $0.022$\,dex. On $104\,799$ DESI stars, $32$ training labels from APOGEE recover the stellar parameters to $129$\,K, $0.25$ and $0.203$\,dex, even for spectra degraded to a signal-to-noise of about $20$. At a signal-to-noise of about $100$, $2048$ labels resolve the two $\alpha$ sequences of the disc in DESI giants, with errors against APOGEE of $51$\,K, $0.07$, $0.047$ and $0.026$\,dex. The code and the data products are at \url{https://github.com/tingyuansen/armillary}.

\end{abstract}

\keywords{Stellar spectral lines (1630), Stellar abundances (1577), Astrostatistics techniques (1886), Dimensionality reduction (1943), Spectroscopy (1558)}

\section{Introduction}
\label{sec:intro}

Spectroscopic surveys now contain tens of millions of stellar spectra \citep{cui2012,majewski2017,gaia2023,desidr1,kollmeier2026}, and the surveys now starting reach fainter stars \citep{dejong2019,weave2024}. Their stellar labels, the effective temperature $\Teff$, the surface gravity $\logg$, the metallicity $\feh$, the $\alpha$-element enhancement $\alphafe$ and the abundances of individual elements, are derived by fitting each spectrum with model spectra \citep[e.g.,][]{garciaperez2016}. Model spectra have known systematic errors, and fitting them to every spectrum of a survey is expensive. The resulting labels disagree between pipelines \citep{jofre2014,smiljanic2014} and between surveys \citep{hegedus2023}, at a level comparable to the astrophysical signals of interest \citep{shen2026}. Placing one survey's spectra on another survey's label scale, without refitting every spectrum from first principles, is therefore a recurring need.

Data-driven models address this need by learning the relation between labels and spectra from a training set of stars and applying it to the rest. The Cannon does so with a polynomial in the labels at every pixel \citep{ness2015,casey2016}. The Payne \citep{ting2019} emulates a grid of model spectra with a neural network, and DD-Payne \citep{ting2017lamost,xiang2019} trains that network on observed stars with the model spectra as a constraint. Supervised neural networks learn the map from spectrum to labels directly \citep{fabbro2018,leungbovy2019,zhang2019starnet}. Authors use these models to transfer labels from one survey to another through the stars that the surveys share \citep[e.g.,][]{ho2017,xiang2019,das2025}.

In all of these the labels enter with the spectra, and the model learns their relation only where labels exist. The training set must therefore span the label space \citep{ness2015}. The more flexible the model, the larger the set it needs, since a direct map from spectrum to labels can also learn correlations that hold in the training set but not in the physics. Such a model gets nothing from the unlabelled majority of a sample, and a star outside its label-space coverage is labelled by extrapolation. Label transfers between two surveys used large training sets of stars in common \citep{ho2017,thomas2024,das2025}.

\begin{figure*}[t]
\centering
\includegraphics[width=\textwidth]{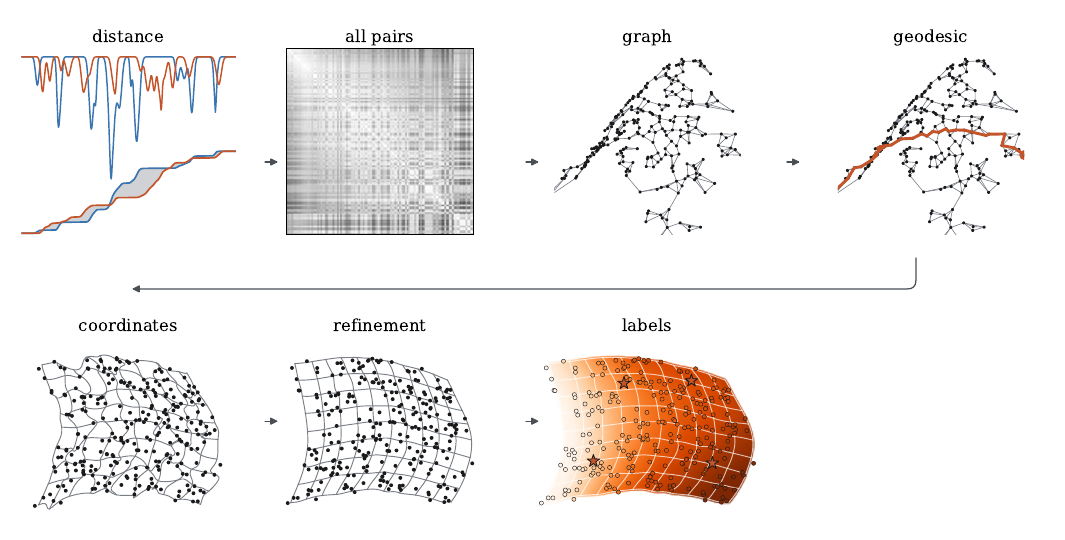}
\caption{A schematic of the steps of Armillary, shown for points on a bent sheet. Top row, from left. Two spectra are compared by the Wasserstein distance (see Fig.~\ref{fig:ruler}) between their signed absorption profiles, the shaded area between their cumulative curves. In this schematic the distance is computed between every pair of stars. A graph joins each star to its nearest neighbours under that distance. The geodesic distance between two distant stars is the length of the shortest path along the graph (orange). Bottom row, the stars in the recovered coordinates, with a regular grid of the sheet drawn through them to show the shape of the manifold (see Fig.~\ref{fig:lattice}). The geodesic distances between all pairs are turned into coordinates, which follow the large-scale shape but scatter on small scales and crumple the grid. A local refinement reduces the scatter by penalising departures from the weighted neighbour relations. The labels of a few labelled stars (orange stars) then determine those of every other star through the same weights, here one label shaded from light to dark.}
\label{fig:schematic}
\end{figure*}

Representation learning exploits the structure of the data before any label is required \citep{bengio2013}. A spectrum is determined by a few parameters, so a set of spectra lies on a low-dimensional manifold in the space of pixel values. The parameters are coordinates on that manifold. Representation learning recovers the manifold from all the spectra, labelled or not. The labels are introduced afterwards, attached to a small training set from which every other star's labels are transferred, which in the language of deep learning is a few-shot use of labels \citep{snell2017}. The unlabelled majority then contributes to the coordinates, and the training set need only cover the part of the manifold it labels. A star far from every training star on the manifold appears so in the coordinates.

Most current representation learning of spectra uses deep learning. Neural networks trained to compress spectra, or segments of them, through a bottleneck, known as autoencoders, return a latent representation of every spectrum \citep{yang2015,sedaghat2021}. Such representations can be expressive, but they are often entangled, with several physical parameters mixed in each latent variable \citep{bengio2013,laroche2025}. Disentangling them requires a supervised model of its own \citep{locatello2018,demijolla2021}, and that model requires the labels the few-shot setting was intended to avoid. A latent space learned without labels can still hold a physical structure, as the two $\alpha$ sequences do for cool giants in one learned from Gaia BP/RP spectra \citep{laroche2025}. Its latent variables, however, have no definition of their own.

The physics of the problem is simpler than that of most data to which deep neural networks are applied. The manifold has a few principal dimensions, the stellar parameters $\Teff$, $\logg$ and $\feh$ and the $\alpha$ enhancement, along which the spectra vary most. Recovering coordinates on a manifold of a few dimensions is a classical problem. We ask whether classical techniques recover them.

The Sequencer of \citet{baron2021} orders a data set along the single dimension in which its objects change most gradually, without labels. Applied to spectra, it ordered stars by temperature and quasars by redshift. Among the distances it uses is one that measures where absorption lies along the wavelength axis, independently of the total amount of absorption. A graph of the pairwise distances is built, and the ordering follows the minimum spanning tree \citep{kruskal1956} of that graph. An ordering, however, assigns a single coordinate to each object, whereas the manifold of stellar spectra has more than one dimension. A single ordering can follow only one of them.

We generalise the Sequencer from one dimension to as many as the spectra vary along. The distance and the graph are retained. The ordering is replaced by coordinates, obtained from the lengths of the shortest paths through the graph, and the number of coordinates is read from the data. We call the method Armillary, after the instrument whose rings are a coordinate system for the sky.

Section~\ref{sec:method} presents Armillary and the reasoning behind each of its steps, and shows that a regular grid of labels is recovered from synthetic spectra. Sections~\ref{sec:apogee} and \ref{sec:surveys} apply it to APOGEE and to DESI. Section~\ref{sec:discussion} relates Armillary to earlier work and states its limitations, and Section~\ref{sec:conclusion} concludes.

\section{Construction of the Coordinates}
\label{sec:method}

Armillary constructs the coordinates of the manifold in four steps. A distance between spectra is defined, and a graph of nearest neighbours is built from it. Coordinates are extracted from the shortest paths through the graph, and a local refinement reduces the noise the graph leaves in them. Figure~\ref{fig:schematic} summarises the steps.

\begin{figure*}[t]
\centering
\includegraphics[width=\textwidth]{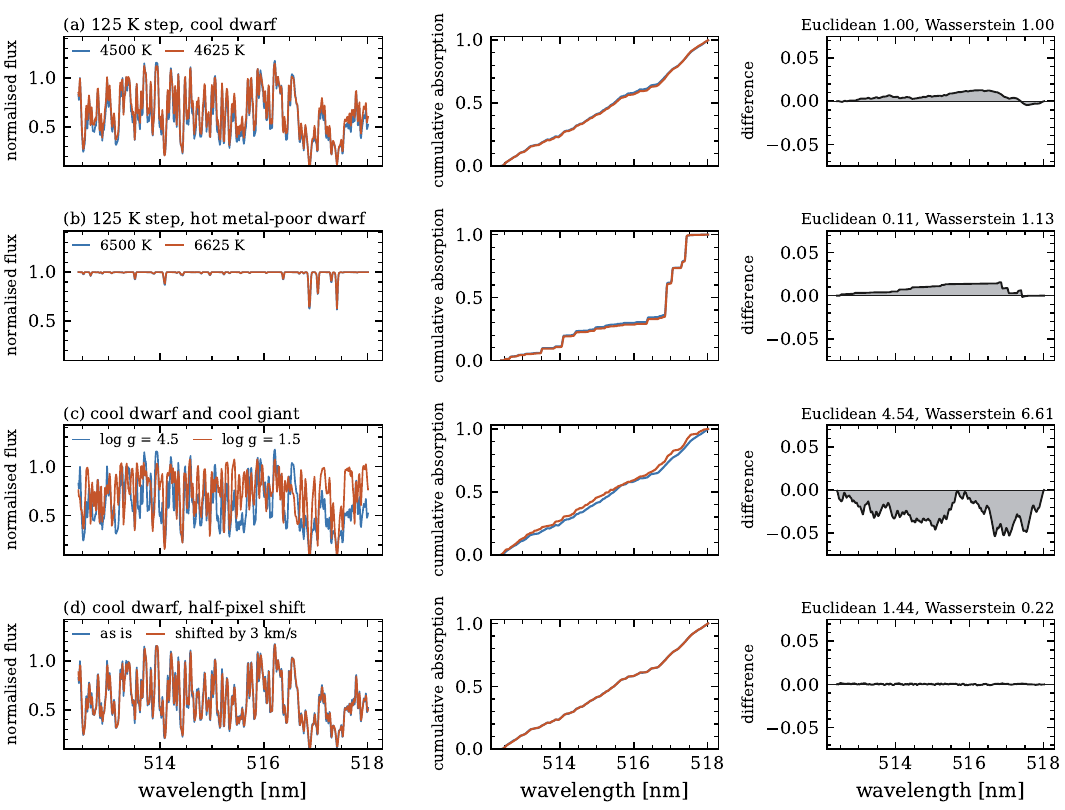}
\caption{The Wasserstein distance measures a change of the spectrum more evenly than the Euclidean distance does, shown on four pairs of synthetic spectra, one pair per row, in the chunk containing the Mg\,b triplet. Left, the normalised flux. Middle, the cumulative absorption curves $F$. Right, their difference, whose integral (shaded) is the Wasserstein distance of the chunk. Above each right-hand panel are the Euclidean and Wasserstein distances between the two complete spectra, over all chunks, each in units of its own value in row (a). Every row then reads as a number of temperature steps of the cool dwarf under each distance. (a) A $125$\,K step for a cool dwarf ($4500$\,K, $\logg=4.5$, solar metallicity). (b) The same step for a hot metal-poor dwarf ($6500$\,K, $\logg=4.5$, $\feh=-2$). (c) The cool dwarf and a cool giant ($\logg=1.5$) of the same temperature and metallicity. (d) The cool dwarf and itself shifted by half a pixel, $3$\,km\,s$^{-1}$ at this resolution.}
\label{fig:ruler}
\end{figure*}

\subsection{The distance between two spectra}
\label{sec:metric}

The distances set the neighbourhoods and the initial coordinates. The refinement also uses the local variation of the normalised spectra. A given change in the parameters must therefore correspond to the same distance wherever it occurs on the manifold. Any unevenness in the distance is inherited by the graph and by the coordinates. The simplest distance, the Euclidean distance between the two spectra as pixel vectors, does not satisfy this requirement, because it is dominated by how much absorption each spectrum contains. A small step in temperature, for example, changes a cool metal-rich spectrum, which is rich in strong lines, by a large amount. It changes a hot metal-poor spectrum, with few lines, by a small one. The coordinates built on such a distance are compressed in the metal-poor corner of the label space and stretched in the metal-rich one.

The Sequencer compares its objects under several distances, each applied to contiguous segments of the data at several scales. Its default set is the Euclidean distance, the Kullback--Leibler divergence \citep{kullback1951}, the Earth Mover's Distance and the energy distance \citep{szekely2013}. \citet{baron2021} note that the Earth Mover's Distance and the energy distance are sensitive to the magnitude of displacements along the pixel axis, whereas the Euclidean distance and the Kullback--Leibler divergence treat each pixel as a separate dimension. We adopt the Wasserstein distance of optimal transport \citep{rubner2000,peyre2018}, which the Sequencer calls the Earth Mover's Distance. As in the Sequencer, we use the distance in its cumulative form and compute it on chunks of the spectrum. We extend it to signed absorption depths so that noise above the continuum need not be clipped. The resulting distance compares the pattern of absorption after the net amount of absorption is normalised. The Euclidean distance serves below as the baseline. Another distance can be substituted where the data call for one.

We divide the wavelength range into $N_{\rm chunk}$ chunks, each a few to a few tens of nanometres wide. Within chunk $c$, star $i$ has an absorption depth $1-f_i(\lambda)$ at each wavelength $\lambda$. Here $f_i$ is the flux divided by a local continuum. Within a window, the continuum is a straight line fitted by least squares to the $15$ per cent of its pixels with the highest flux. The window width is set for each survey. The depths define a signed absorption profile along the wavelength axis.

Dividing the depth by the net absorption in the chunk gives the normalised profile $\rho_i^{(c)}$. Its integral is one, whatever the strength of the lines, provided the net absorption is nonzero. A pixel where noise puts the flux above the continuum has a negative depth, and the depth is kept negative. Noise scatters the flux above and below the continuum by equal amounts, so on average it adds no net absorption.

The depth is also weighted where the pixel errors carry structure of their own, from the sky and the detector. Each pixel's depth is multiplied by a reliability weight $v_i(\lambda)=\sigma_i^{-2}(\lambda)/\langle\sigma_i^{-2}\rangle_c$, the pixel's inverse variance divided by its mean over the good pixels of the chunk. A noisy pixel then carries less of the chunk's absorption, and the rest of the chunk keeps its weight. Where the errors carry no such structure, $v_i=1$. Integrating $\rho_i^{(c)}$ from the start of the chunk $\lambda_c$ gives the cumulative curve $F_i^{(c)}$,
\begin{align}
\rho_i^{(c)}(\lambda)&=\frac{v_i(\lambda)\,[1-f_i(\lambda)]}{\int_c v_i(\lambda')\,[1-f_i(\lambda')]\,\dd\lambda'},
\label{eq:rho}\\
F_i^{(c)}(\lambda)&=\int_{\lambda_c}^{\lambda}\rho_i^{(c)}(\lambda')\,\dd\lambda' .
\label{eq:curve}
\end{align}
The superscript names the chunk and the subscript the star. For positive net absorption the cumulative curve rises through absorption lines and can fall where the flux lies above the continuum. In the pixel implementation, a net depth sum below $10^{-3}$ in absolute value, in practice only that of a fully masked chunk, is replaced by $10^{-3}$ in the denominator.

For nonnegative unit-normalised profiles, the one-dimensional Wasserstein distance, written $\Wone$, is the minimum cost of moving absorption from one profile to the other, with cost equal to the amount of absorption moved times the wavelength distance it is moved. The subscript is the power to which that distance is raised in the cost, here one. The distance equals the area between the cumulative curves of the two profiles, and we keep that expression for the signed profiles,
\begin{equation}
\Wone^{(c)}(i,j)=\int_c\big|F_i^{(c)}(\lambda)-F_j^{(c)}(\lambda)\big|\,\dd\lambda ,
\label{eq:w1}
\end{equation}
shown in Figure~\ref{fig:ruler}. With negative depths the profiles are signed and $F$ need not be a cumulative distribution function. The expression remains the $L^1$ distance between the two cumulative profiles. It extends the Wasserstein distance to signed profiles with the same integral, and we keep that name throughout.

The chunk distances are then combined into one. Summed directly, they would be dominated by the widest chunks, because a wide chunk moves absorption over longer distances. We therefore divide each chunk's distance by its median over pairs of stars, the typical distance in that chunk, and sum over the chunks,
\begin{equation}
D_{ij}=\sum_{c=1}^{N_{\rm chunk}}\frac{\Wone^{(c)}(i,j)}{\operatorname{median}_{(i',j')}\Wone^{(c)}(i',j')} ,
\label{eq:metric}
\end{equation}
so that every chunk contributes in the same units, the number of typical distances by which the two stars differ in that chunk. The wavelength range can be divided into chunks in more than one way, into a few wide chunks and into many narrow ones, and the sum then runs over the chunks of every division. A wide chunk compares the broad features of two spectra and a narrow one their individual lines.

Figure~\ref{fig:ruler} compares the Euclidean and Wasserstein distances between complete spectra in three cases, with one chunk of each pair drawn. The spectra are synthetic, computed with Payne-Zero \citep{tingkim2026}, a Python implementation of Kurucz's ATLAS12 and SYNTHE \citep{kurucz2005} with the synthesis on GPUs, at a resolving power of $20\,000$ over $480$--$680$\,nm without noise. They are divided into $32$ chunks each about $6$\,nm wide.

The first case, rows (a) and (b), is a small step in temperature for two stars. One is a cool dwarf of solar metallicity, whose spectrum is crowded with strong lines, and the other a hot metal-poor dwarf, whose spectrum has few and weak lines. Once the total is divided out, each has one unit of absorption in every chunk, and the Wasserstein distance does not see that the cool dwarf absorbs far more. It measures where along the chunk that unit lies, and the temperature step shifts that location by a similar amount in both stars. Under the Euclidean distance the step is nine times shorter for the hot metal-poor dwarf than for the cool dwarf. Under the Wasserstein distance the two are within fifteen per cent.

The second case, row (c), is the cool dwarf and a cool giant of the same temperature and metallicity. They have similar amounts of absorption but differ in its pattern, for example in the pressure-broadened wings of the strong lines. The Wasserstein distance separates them more widely than the Euclidean distance does, by six and a half steps against four and a half. It does so because the amount of absorption is divided out, and only the pattern remains.

The third case, row (d), is the cool dwarf against itself shifted by half a pixel, as by an uncorrected radial-velocity offset. The Euclidean distance compares the two spectra at matched wavelengths, so every shifted line appears as two residuals of opposite sign. That shift gives a larger distance than a temperature step. The Wasserstein distance responds to a displacement through changes in the cumulative absorption profile, so the same shift gives $0.22$ of a temperature step. The Wasserstein distance is less sensitive to a small residual radial-velocity error, which changes none of the star's parameters.

\subsection{The neighbour graph and geodesic distances}
\label{sec:lattice}

The distance $D$ is intended to measure local changes of a spectrum. It need not remain proportional to parameter separation between distant stars. In the synthetic temperature sequence, for example, it grows more slowly than the number of parameter steps. Nonnegative unit-normalised profiles would impose a chunk-width upper bound on $\Wone^{(c)}$, but the signed profiles used here do not have that bound.

The distance along the manifold between two distant stars is therefore assembled from short distances. We build the neighbour graph $\mathcal{G}$ by finding the $k$ nearest neighbours of every star under $D$. Stars $a$ and $b$ are joined by an edge if $a$ is among the $k$ nearest neighbours of $b$ or $b$ among those of $a$, and the edge has length $D_{ab}$. Such a graph can fall into a few separate pieces, since a group of unusual stars can have all its nearest neighbours within the group. A star in such a piece would have no path to the rest. Every smaller piece is therefore joined to the largest piece by a single edge, the shortest of all the edges that could join the two, and $\mathcal{G}$ is connected. A bridge can span a sparsely sampled region, where $D$ no longer tracks the separation, so the piece's position relative to the rest is only approximate.

The geodesic distance between stars $i$ and $j$ is the length of the shortest path through $\mathcal{G}$ between them, the sum of $D_{ab}$ over the edges $(a,b)$ of the path,
\begin{equation}
D^{\rm g}_{ij}=\min_{\text{paths }i\to j\text{ in }\mathcal{G}}\ \sum_{(a,b)\in\text{path}} D_{ab} ,
\label{eq:geodesic}
\end{equation}
found by Dijkstra's algorithm \citep{dijkstra1959}, which grows a tree of shortest paths outward from $i$ by always extending the closest star not yet reached, and so visits every star once. Apart from the bridges, the path uses neighbour distances, for which $D$ is intended to be most informative. The geodesic distance can therefore keep growing with separation along the manifold where the direct distance $D$ no longer tracks that separation. The Sequencer orders its objects by a walk through the minimum spanning tree of their combined distances, and it is that ordering we replace with coordinates.

\subsection{Coordinates from geodesic distances}
\label{sec:coordinates}

The geodesic distances form a matrix $D^{\rm g}$, $N^2$ numbers between pairs of stars, and coordinates turn them into a position for each star. Classical multidimensional scaling \citep{torgerson1952} gives points $\mathbf{c}_i$ in $d$ dimensions by approximating the inner products implied by $D^{\rm g}$, so that their Euclidean separations approximate the graph geodesics. Exact agreement is possible when the intrinsic distances are Euclidean, as for a bent copy of a convex region of flat space. The coordinates then recover positions in that space.

We first convert the distances into inner products, with the origin at the mean of the stars. An eigendecomposition of the inner products then gives the coordinates. The squared distance between two points is set by their lengths and their inner product,
\begin{equation}
|\mathbf{c}_i-\mathbf{c}_j|^2=|\mathbf{c}_i|^2+|\mathbf{c}_j|^2-2\,\mathbf{c}_i\!\cdot\!\mathbf{c}_j ,
\end{equation}
so the matrix of squared distances is $-2$ times the matrix of inner products plus the squared lengths $|\mathbf{c}_i|^2$ and $|\mathbf{c}_j|^2$. The squared length of star $i$ is the same in every entry of row $i$, and that of star $j$ in every entry of column $j$. Subtracting the row and column means removes both and leaves $-2$ times the inner products, and the factor $-\tfrac12$ then removes the $-2$. The centring matrix $J$ performs the subtraction,
\begin{equation}
B=-\tfrac12\,J\,(D^{\rm g})^{2}\,J,\qquad J=I-\tfrac1N\mathbf{1}\mathbf{1}^{\!\top} ,
\label{eq:mds}
\end{equation}
where $N$ is the number of stars, $I$ the identity, $\mathbf{1}$ the vector of ones, and the square of $D^{\rm g}$ is taken entry by entry. For Euclidean distances the result is the matrix of inner products $B_{ij}=\mathbf{c}_i\!\cdot\!\mathbf{c}_j$ of coordinates whose mean lies at the origin.

In that case, writing the coordinates as a matrix $C_{\rm geo}$ with one row per star gives $B=C_{\rm geo}C_{\rm geo}^{\!\top}$. The eigenvectors of $B$, each scaled by the square root of its eigenvalue, are therefore the columns of $C_{\rm geo}$. Keeping the eigenvectors of the $d$ largest positive eigenvalues gives the $d$ coordinates whose inner products reproduce $B$ as closely as any $d$ coordinates can. Each positive eigenvalue measures how much of $B$ one coordinate accounts for. A nearly Euclidean manifold can give a small set of dominant eigenvalues, which guides how many coordinates to keep.

\subsection{A locally linear refinement}
\label{sec:refinement}

The geodesic coordinates capture the large-scale arrangement but can scatter on smaller scales. A shortest path through the graph depends on which stars sample the route. Detours through those stars can lengthen it, while edges across curved regions can shorten it, and the coordinates inherit these sampling errors. Even a small relative error in a long geodesic distance can disturb a star's placement among its immediate neighbours.

The refinement reduces that scatter by using the local structure of the manifold, following \citet{roweis2000}. On a smooth manifold every point is nearly a weighted average of its nearest neighbours. We take each star's nearest neighbours $\mathcal{N}(i)$ under the distance $D$ between spectra. Let $\mathbf{x}_i$ be the projection of star $i$'s spectrum onto the leading principal components of the sample. The projection retains the dominant spectral variation while reducing the pixel noise entering the local fits. The weights $w_{ij}$, summing to one, are those that best reconstruct $\mathbf{x}_i$ from the projections of its neighbours,
\begin{equation}
\mathbf{x}_i\approx\sum_{j\in\mathcal{N}(i)}w_{ij}\,\mathbf{x}_j,\qquad \sum_j w_{ij}=1 .
\label{eq:lle}
\end{equation}
The weights are found in closed form, as one small linear solve per star. Let $Z_i$ be the matrix whose rows are the differences $\mathbf{x}_j-\mathbf{x}_i$ from the star to its neighbours, and $H_i=Z_iZ_i^{\!\top}$ the matrix of their inner products. The weights solve $[H_i+\epsilon\,\mathrm{tr}(H_i)I]\,\mathbf{w}_i=\mathbf{1}$ with a small ridge $\epsilon$ and are then scaled to sum to one.

The weights describe how star $i$ sits among its neighbours, so they can be imposed on the coordinates. They come from the spectra and $D$ alone, so the noise of the geodesic coordinates does not enter them. We use them to regularise the geodesic coordinates, asking for coordinates that keep their large-scale shape while making every star the same weighted average of its neighbours.

Let $W$ be the matrix with entries $w_{ij}$, one row per star like $C_{\rm geo}$, and $C$ the refined coordinates we seek, arranged the same way. The row $i$ of $(I-W)\,C$ is star $i$'s coordinate minus the weighted average of its neighbours', the amount by which the neighbour relation fails for that star. The two requirements are that every star is the weighted average of its neighbours under the weights $w_{ij}$, and that $C$ stays close to $C_{\rm geo}$. The coordinates that balance them are
\begin{equation}
C=\arg\min_{C}\ \big\|(I-W)\,C\big\|^2+\mu_{\rm c}\,\bar{m}\,\big\|C-C_{\rm geo}\big\|^2 ,
\label{eq:refine}
\end{equation}
in which the norms are sums of squares over every entry, and $\bar{m}$ is the mean diagonal entry of $(I-W)^{\!\top}(I-W)$, which sets the scale of the first term. The minimum is taken over every $N\times d$ matrix $C$, and $\arg\min$ returns the one that attains it. The first term enforces the neighbour relations, the second holds the coordinates near $C_{\rm geo}$, and the anchor strength $\mu_{\rm c}$ sets the trade-off between them. The quantity minimised is quadratic in $C$, so its minimum is where its gradient with respect to $C$ vanishes. That gradient is $2(I-W)^{\!\top}(I-W)\,C+2\mu_{\rm c}\,\bar{m}\,(C-C_{\rm geo})$, and setting it to zero gives one linear solve for $C$,
\begin{equation}
\big[(I-W)^{\!\top}(I-W)+\mu_{\rm c}\,\bar{m}\,I\big]\,C=\mu_{\rm c}\,\bar{m}\,C_{\rm geo} .
\label{eq:solve}
\end{equation}
As $\mu_{\rm c}\to\infty$ the solution is $C_{\rm geo}$ itself. At smaller $\mu_{\rm c}$, coordinate variations that poorly satisfy the neighbour relations are suppressed more strongly, and this can include physical structure as well as noise. The geodesic anchor retains the global arrangement, and the recovered synthetic grid provides a check on the retained structure.

\subsection{Transferring labels through the graph}
\label{sec:reading}

For fixed settings, the coordinates, graph and refinement weights are built from the spectra of all stars, and no label enters them. Most settings were chosen with labelled stars in a separate sweep (Section~\ref{sec:apogee_coords}), and these labels are not counted in the training set. Only the labels of the training set enter the label transfer, and every other star receives its labels from them through the refinement weights.

The problem has the form of the refinement with two differences. The labels take the place of the coordinates, and the anchor acts only on the training set, pulling each of its stars toward its known labels, where the refinement pulled every star toward $C_{\rm geo}$. Let $\mathbf{y}_i$ be the known labels of star $i$ and $Y$ the labels of every star, one row per star. The labels are
\begin{equation}
Y=\arg\min_{Y}\ \big\|(I-W)\,Y\big\|^2+\mu_{\rm y}\,\bar{m}\sum_{i\in\mathcal{T}}\big\|Y_i-\mathbf{y}_i\big\|^2 ,
\label{eq:propagate}
\end{equation}
where the sum runs over the training set $\mathcal{T}$, $Y_i$ is row $i$ of $Y$, and the anchor strength $\mu_{\rm y}$ sets how strongly the training set is pulled toward its labels. Equation~\ref{eq:propagate} is read as Equation~\ref{eq:refine}, with the minimum taken over the labels $Y$ of every star. As for the refinement, setting the gradient with respect to $Y$ to zero gives one linear solve,
\begin{equation}
\big[(I-W)^{\!\top}(I-W)+\mu_{\rm y}\,\bar{m}\,P\big]\,Y=\mu_{\rm y}\,\bar{m}\,P\,\hat{Y} ,
\label{eq:propsolve}
\end{equation}
where $P$ is the diagonal matrix with ones on the rows of the training set and zeros elsewhere, and $\hat{Y}$ holds the known labels on those rows and zeros elsewhere. The labels of all stars are determined together, so a star far from the training set receives its labels through the stars between it and the nearest training star. The label transfer adds one parameter, the anchor strength $\mu_{\rm y}$.

\subsection{Implementation}
\label{sec:scaling_method}

A direct implementation would form several dense matrices over the stars. The matrix of $D$ takes memory proportional to $N^2$, about $160$\,GB in single precision for $N=2\times10^{5}$. Dijkstra's algorithm from all $N$ stars on the sparse $k$-neighbour graph costs $O(kN^2\log N)$, or $O(N^2\log N)$ for fixed $k$, and storing all geodesic distances also takes $O(N^2)$ memory. The dense eigendecomposition of $B$ and the dense refinement solve cost $O(N^3)$. We therefore replace these steps by computations that avoid the full dense matrices and give the same result, exactly or to a close approximation.

Only each star's nearest neighbours enter the graph, so the matrix of $D$ is never needed, only a search for those neighbours. The integral that gives the chunk distance is a sum of absolute differences over the pixels, so each star can be written as one vector of its cumulative curves scaled chunk by chunk. $D$ is then the sum of absolute differences between two such vectors, the form that nearest-neighbour search handles. The median that scales each chunk is taken over the pairs among a few hundred stars drawn at random. For the $2\times10^{5}$ APOGEE stars, the lists of $100$ neighbours and their distances take $0.16$\,GB in place of $160$\,GB. The largest array is then the vectors themselves, $18$\,GB.

The neighbours are found by nearest-neighbour descent \citep{dong2011}, which starts from random neighbour lists and improves them by testing the neighbours of each star's current neighbours. Its computational cost depends on the sample geometry and the number of iterations, and it avoids evaluating every pair. For $256$ stars drawn at random in each survey, the $100$-neighbour lists overlap those from an exact search by more than $99.8$ per cent on average.

The shortest paths and the eigendecomposition are run through a subset of stars, following \citet{desilva2004}. A few hundred stars, $M$ of them, are drawn at random as landmarks. Dijkstra's algorithm runs from those $M$ stars alone, reducing the shortest-path cost to $O(MkN\log N)$, and the eigendecomposition is applied to the $M\times M$ matrix of geodesic distances among them. Every other star is then placed from its geodesic distances to the landmarks, as a position is fixed from its distances to a few known points. Its squared distances, less the landmarks' mean squared distances among themselves, are the star's row of the centred matrix of squared distances. The inner products with the landmark coordinates that follow from it, divided by the eigenvalues, are its coordinates.

The refinement and the label transfer are sparse linear systems, since a row of $W$ has only as many entries as a star has neighbours. They are solved by conjugate gradients \citep{hestenes1952}, which does not form or invert the matrix. It needs only the matrix's product with a vector, two passes over the entries of $W$, repeated until the residual falls below a set tolerance.

Each replacement was checked against the dense original on a subset small enough to run both. The whole chain from spectra to coordinates takes $45$ minutes for the $2\times10^{5}$ APOGEE stars, $60$ per cent of it in the neighbour search. It used $24$ threads on one node of the Ohio Supercomputer Center's Pitzer cluster, with two $24$-core Intel Xeon Platinum 8268 processors at $2.9$\,GHz.

\begin{figure*}[t]
\centering
\includegraphics[width=\textwidth]{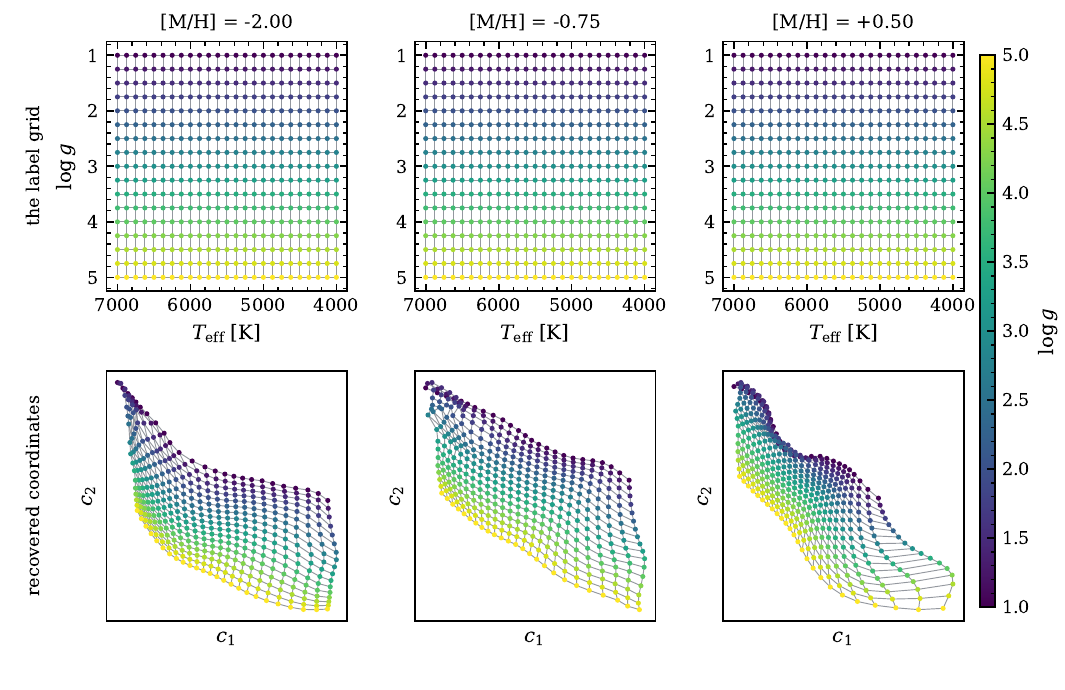}
\caption{Armillary applied to the synthetic spectra of a $25\times17\times11$ grid in $\Teff$, $\logg$ and $\feh$ (spectra as in Fig.~\ref{fig:ruler}, steps as in Fig.~\ref{fig:schematic}). Each column is the grid's stars at one of its eleven metallicities. Top, those stars in $\Teff$ and $\logg$. Bottom, the same stars in the first two recovered coordinates $c_1$ and $c_2$, with lines joining neighbouring grid points. The three metallicities, and lines that cross in this projection, are separated along $c_3$, which is not drawn. Colour is $\logg$ in both rows. The grid comes back with its lines in order.}
\label{fig:lattice}
\end{figure*}

\subsection{Recovering a regular label grid from synthetic spectra}
\label{sec:synthetic}

The aim of the construction is coordinates that are the stellar labels up to a smooth transformation. Real spectra cannot test this directly, because their labels are themselves estimates. Synthetic spectra can, since the labels that generated them are known exactly, and a regular grid of labels makes the test visual.

We build a grid of $25\times17\times11$ stars. $\Teff$ runs from $4000$ to $7000$\,K in steps of $125$\,K, $\logg$ from $1$ to $5$ in steps of $0.25$, and $\feh$ from $-2$ to $+0.5$ in steps of $0.25$. Their spectra are synthesised with Payne-Zero as in Figure~\ref{fig:ruler}, and Armillary receives only the spectra. The settings are those adopted for APOGEE and DESI, adapted to the small grid, with the distance using the same $32$ chunks as Figure~\ref{fig:ruler} and all $4675$ stars used as landmarks. If its coordinates are the labels up to a smooth transformation, the grid comes back as a smoothly deformed grid with its lines in order. Otherwise the grid folds.

Figure~\ref{fig:lattice} shows the outcome at three of the eleven metallicities. The eigenvalues single out three coordinates, one per label, with the fourth eigenvalue $3.3$ times smaller than the third. In every slice the grid comes back with temperature running one way and gravity the other. Lines can cross in the projection onto $c_1$ and $c_2$, but the grid does not fold in the three coordinates.

The recovered grid is not a rectangle, since the coordinates record how much the spectra change from one grid point to the next and that change differs across the grid. A step of $125$\,K moves a cool spectrum further than a hot one. This is because many lines and the molecular bands respond to temperature at the cool end, while few lines remain at the hot end. The recovered grid is therefore stretched at its cool end and compressed at its hot end. A temperature step is $1.6$ times longer at $\Teff\le4500$\,K than at $\Teff\ge6500$\,K, and twice as long for the cool metal-rich dwarfs, whose crowded spectra respond most. A step in metallicity is $1.4$ times longer at the metal-poor end than at the metal-rich end, because unsaturated lines grow in proportion to the abundance and saturated ones barely grow \citep{gray2005}. Gravity steps are nearly even.

Under the Euclidean distance the same effect is larger, a factor of $41$ between the largest and smallest median grid steps in the four combinations of cool or hot and metal-poor or metal-rich stars. That distance also counts how much absorption each spectrum has. The Wasserstein distance divides the amount of absorption out and reduces the factor to $4$. The grid is still not even, but the factor is ten times smaller than under the Euclidean distance. For temperature steps alone, the longest is $2.3$ times the shortest, against $9$ times under the Euclidean distance.

\section{Application to APOGEE Spectra}
\label{sec:apogee}

APOGEE \citep{majewski2017,wilson2019} is a near-infrared survey at $1.51$--$1.70\,\mu$m and resolving power $22\,500$. Its pipeline ASPCAP \citep{garciaperez2016,jonsson2020} fits each spectrum with model spectra and reports $\Teff$, $\logg$, the overall metallicity $\feh$, $\alphafe$, $\cm$ and $\nm$ from the fit of the whole spectrum. Individual abundances come from windows on their lines. The first four are the reference labels here. From DR17 \citep{abdurrouf2022} we take every main-survey star with no ASPCAP bad bit, no star-level flag, finite labels and a signal-to-noise above $70$. This gives $202\,932$ stars. Each enters with its combined spectrum from the reduction pipeline \citep{nidever2015}, the one of highest signal-to-noise where a star has more than one.

We use the pipeline's pseudo-continuum-normalised spectra, which sit on three detectors with gaps between them. Chunks are built per detector so that none straddles a gap, in three chunkings of $11$, $20$ and $40$ chunks, $16$ to $4$\,nm wide. The local continuum is fitted in the chunks of the middle chunking, about eight nanometres wide, on top of the pipeline's normalisation. A slow tilt of that normalisation across a chunk would otherwise shift the absorption distribution as a real feature does. A pixel the pipeline masks has weight zero in the cumulative curve.

\subsection{The coordinates as physical axes}
\label{sec:apogee_coords}

Armillary is applied to the $202\,932$ spectra as follows. We join each star to its $k=30$ nearest neighbours and draw $M=800$ landmarks at random. We compute the weights of Equation~\ref{eq:lle} from each star's $100$ nearest neighbours under the Wasserstein distance $D$, on a projection of $30$ components with a ridge of $10^{-3}$. The anchor strengths are $\mu_{\rm c}=0.003$ for the refinement (Equation~\ref{eq:refine}) and $\mu_{\rm y}=3$ for the label transfer (Equation~\ref{eq:propagate}). Apart from $\mu_{\rm c}$, the settings come from the sweep below. We chose $\mu_{\rm c}$ without labels, on the $11\,252$ stars that APOGEE observed twice, whose two spectra must receive the same coordinates. Their separation falls as $\mu_{\rm c}$ decreases and flattens below $0.003$, which is adopted.

\begin{figure*}[t]
\centering
\includegraphics[width=\textwidth]{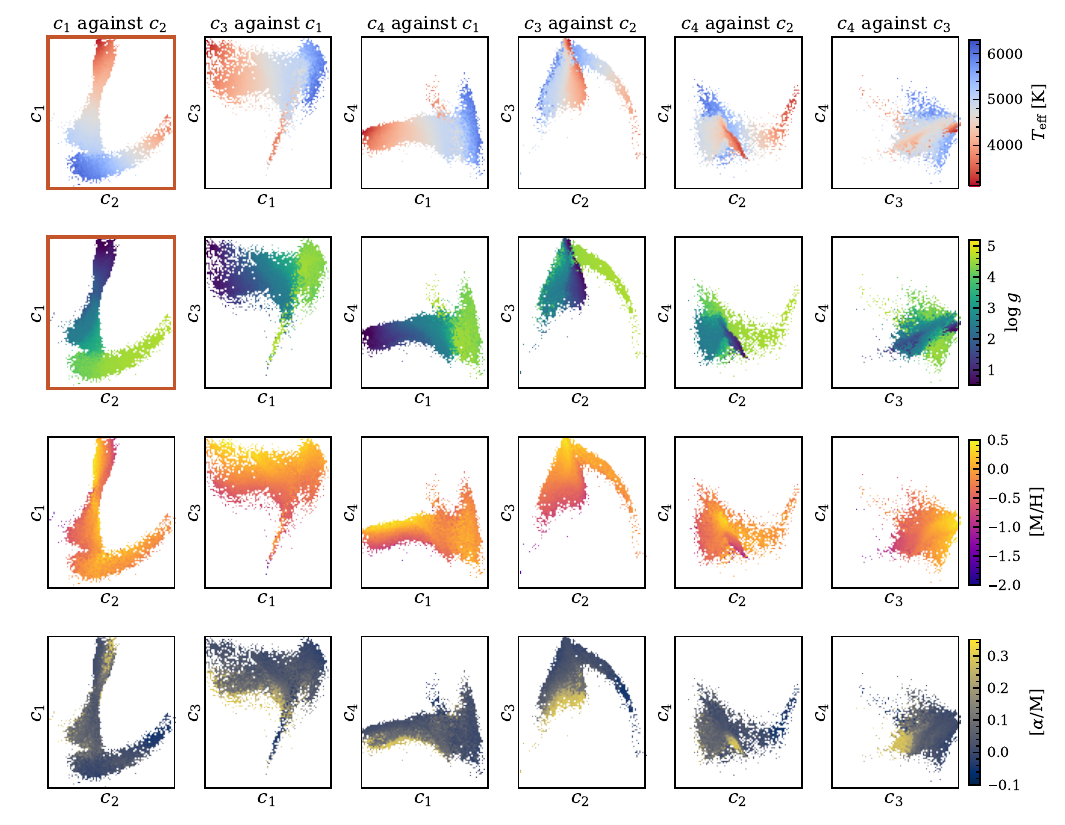}
\caption{Armillary's four coordinates $c_1$ to $c_4$ of the $202\,932$ APOGEE stars. Each column plots one pair of coordinates, named above it as $y$ against $x$ in arbitrary units. Each row colours the same points by one ASPCAP label, $\Teff$, $\logg$, $\feh$ or $\alphafe$, none of which entered the coordinates. Each cell shows the median label of the stars in it and is drawn where at least three fall. The boxed panels are the first two coordinates coloured by $\Teff$ and by $\logg$. They already resemble the Kiel diagram, the giant branch rising to the upper right and the main sequence along the bottom, although no label was used in placing a star. Metallicity runs along $c_3$.}
\label{fig:coordinates}
\end{figure*}

\begin{figure*}[t]
\centering
\includegraphics[width=\textwidth]{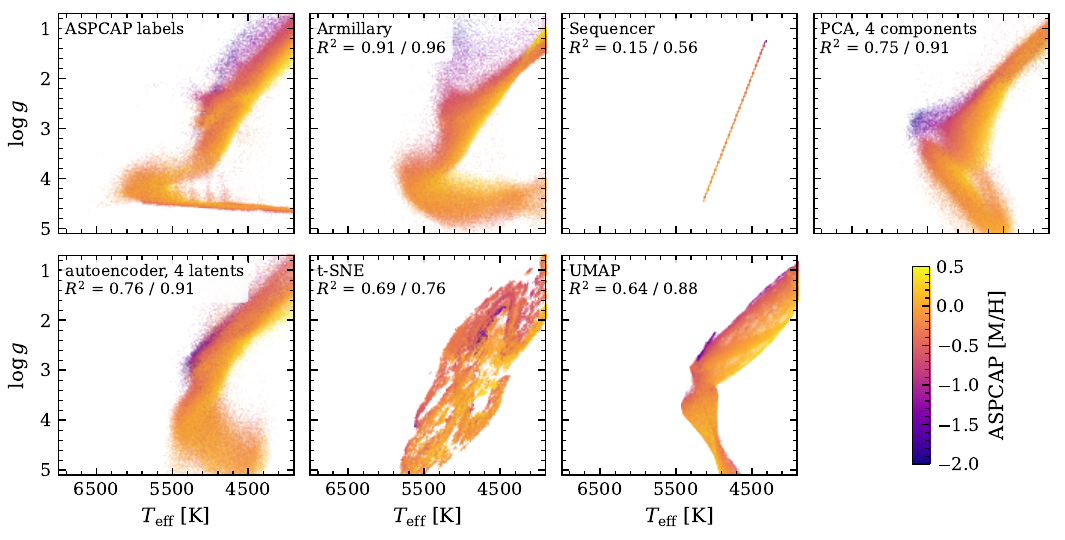}
\caption{The Kiel diagram of the $202\,932$ APOGEE stars in ASPCAP's labels (top left), and the same stars placed by the linear probe, the best linear map from each representation's output to $(\Teff,\logg)$. Each panel gives the $R^2$ of the map, the fraction of the variance of $\Teff$ and of $\logg$ it accounts for. Colour is ASPCAP $\feh$ throughout. Armillary, PCA, UMAP and the autoencoder are four-dimensional, t-SNE three, and the Sequencer ordering is one number per star, so its map is a line.}
\label{fig:competition}
\end{figure*}

\begin{figure*}[t]
\centering
\includegraphics[width=\textwidth]{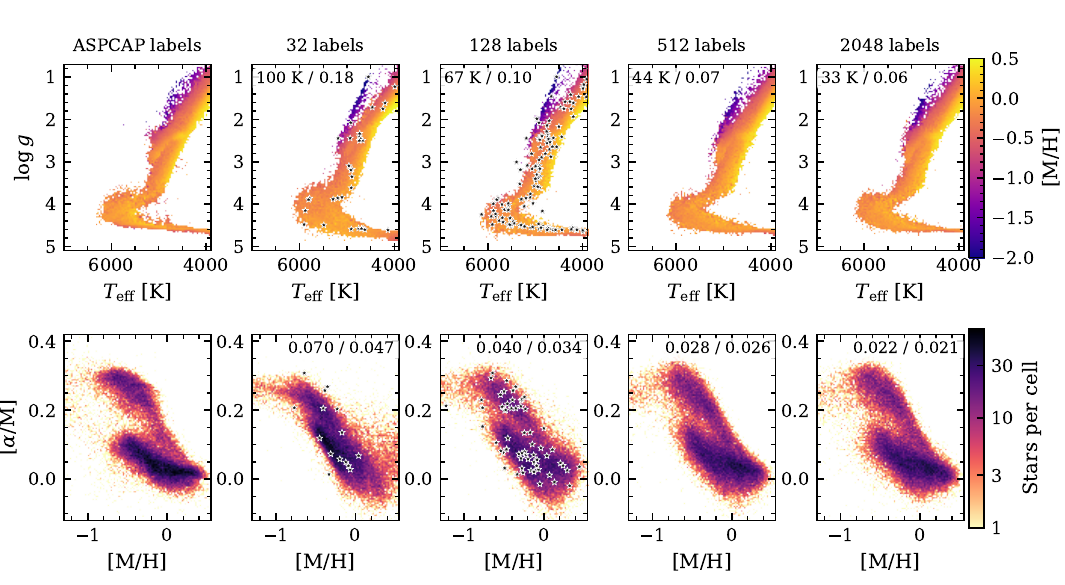}
\caption{The labels of a training set transferred to the $202\,932$ APOGEE stars through Armillary's graph. Top, the Kiel diagram coloured by $\feh$, and bottom, the $\alphafe$--$\feh$ plane of the giants ($\logg<3.5$) as a density on a logarithmic scale. The first column is ASPCAP's labels. The next four are the labels transferred from training sets of $32$, $128$, $512$ and $2048$ stars, every axis and the colour being the transferred labels. The star symbols on the second and third columns mark the $32$ and $128$ training stars at their ASPCAP labels, the giants among them below. The numbers are the $1\sigma$ errors against ASPCAP, $\Teff$ and $\logg$ above over every star not in the training set, and $\feh$ and $\alphafe$ below over the giants shown.}
\label{fig:showcase}
\end{figure*}

\begin{figure*}[t]
\centering
\includegraphics[width=\textwidth]{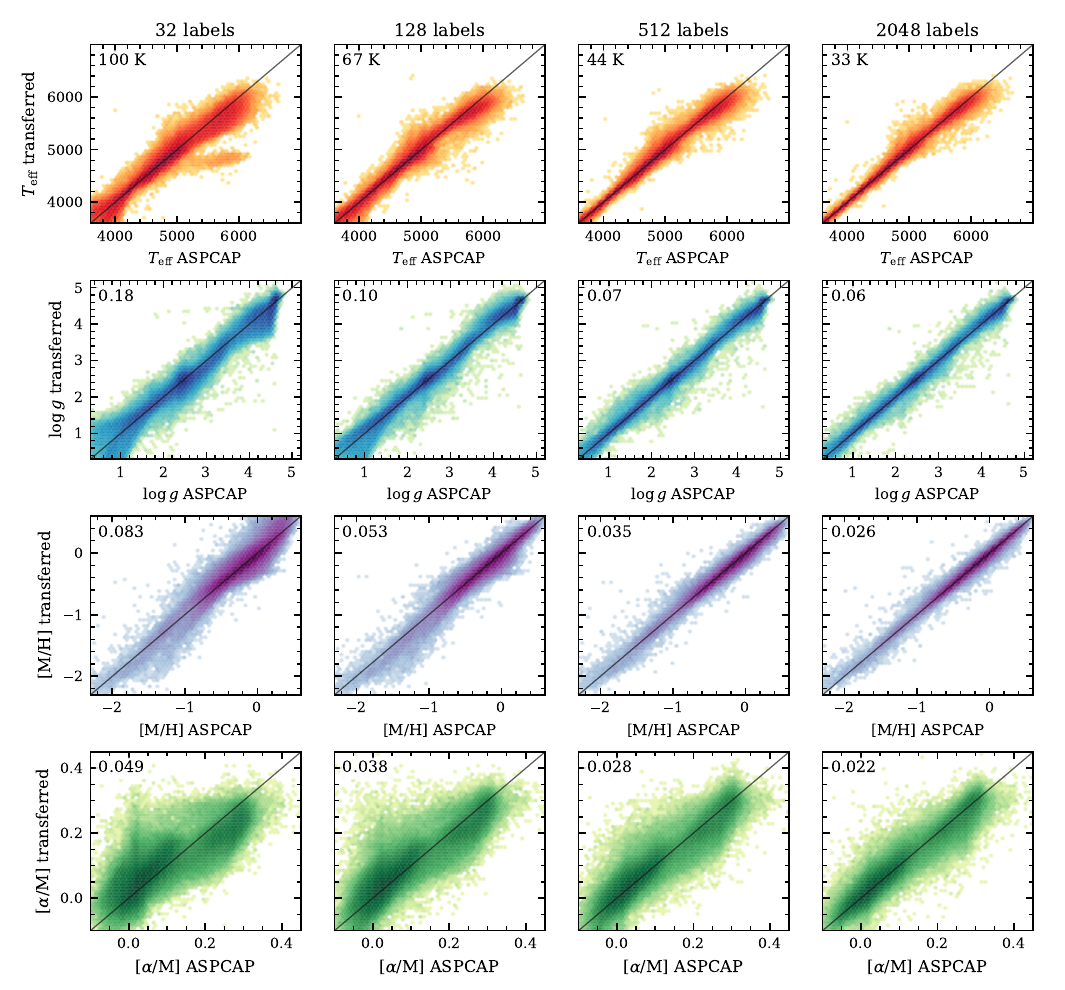}
\caption{The labels transferred to the APOGEE stars against ASPCAP's, one row per label and one column per training set of $32$, $128$, $512$ and $2048$ stars, on every star not in the training set. Each panel is the density of the stars on a logarithmic scale, with the one-to-one line, and its number is the $1\sigma$ error of the difference. All four rows use the same held-out stars, including dwarfs and giants.}
\label{fig:onetoone_apogee}
\end{figure*}

\begin{figure*}[t]
\centering
\includegraphics[width=\textwidth]{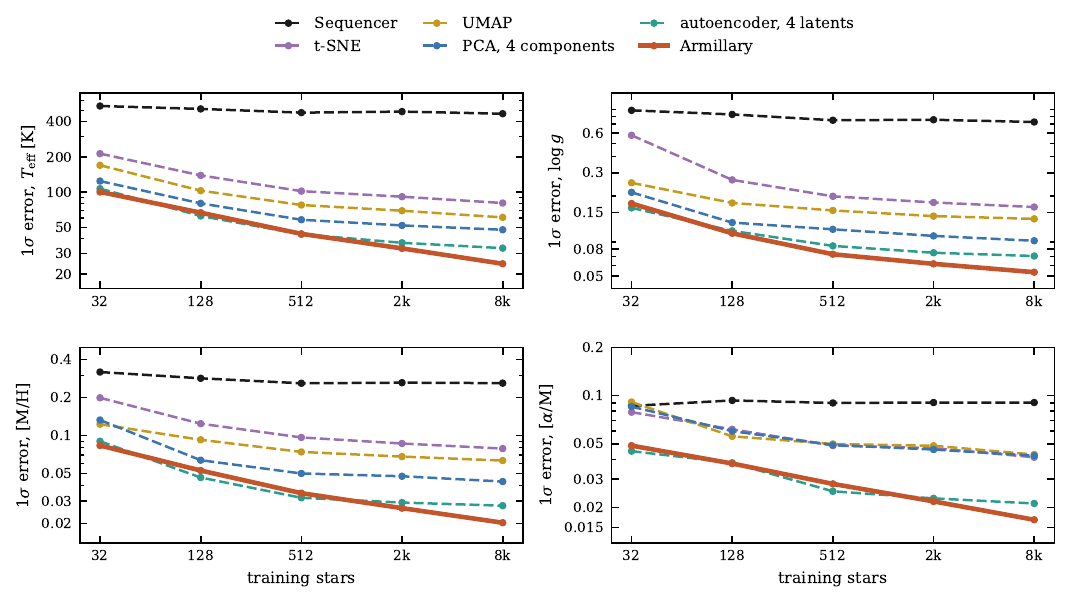}
\caption{The $1\sigma$ error of each label against the size of the training set, on the $202\,932$ APOGEE stars, evaluated on every star outside each training set. The same training set is given to every representation. Armillary transfers its labels through the graph (orange, solid), and every other representation (dashed) by a multilayer perceptron from its output to the labels. Each perceptron is fitted on the training set at the weight decay of lowest error, chosen separately for each of the four labels.}
\label{fig:ladders}
\end{figure*}

The other settings were fixed by one sweep on a random $60\,000$ of the stars. Each setting was varied in turn about the values above. Labels were transferred through the graph from training sets of $64$, $256$, $1024$ and $16\,384$ stars, and the errors of the four labels were measured on held-out labelled giants. The number of neighbours of the weights was swept from $50$ to $1600$, $\mu_{\rm y}$ from $0.3$ to $30$, and the projection from $10$ components to the spectrum itself. The ridge $\epsilon$ of the weights' solve was swept over a factor of a hundred, and $10^{-3}$ was adopted.

Averaging relative errors over labels and training sizes, $100$ neighbours give the minimum, with the error $15$ per cent higher at $50$ and $7$ per cent higher at $400$. Projections of $20$ and $30$ components are within $4$ per cent of each other and $30$ is adopted. The error decreases as $\mu_{\rm y}$ rises to $3$ and changes by less than $4$ per cent for larger values. The ridge of $10^{-3}$ is the best for training sets of $256$ stars and fewer. Smaller ridges lower the $\alphafe$ error by $15$--$16$ per cent at $1024$ training stars and $3$--$5$ per cent at $16\,384$. The linear-probe $R^2$ of the coordinates changes by at most $0.05$ in any label for graphs of $10$ to $100$ neighbours and $400$ to $1600$ landmarks. The paper's three scales of chunks give an error $1$ to $3$ per cent lower than one, two or four scales.

Four coordinates are kept to represent the three stellar parameters and the $\alpha$ enhancement. The leading eigenvalues of $B$ describe how the embedding is distributed across its dimensions. They fall as $1$, $0.35$, $0.14$, $0.13$, $0.09$, $0.07$, the fifth $30$ per cent below the fourth. This is not a sharp separation of four physical axes, since weaker dimensions can contain noise and further spectral variation from individual abundances, and the geodesic distances need not be exactly Euclidean. Keeping four affects the coordinates and the draw of training stars (Section~\ref{sec:showcase}), but not the label transfer, whose weights come from the spectra. Where no gap is clear, more coordinates can be kept at little cost, since the eigendecomposition returns them all.

Figure~\ref{fig:coordinates} shows every pair of the four coordinates, coloured afterwards by each label. The first coordinate runs along the Kiel diagram, from hot dwarfs to cool giants, the second across it, separating the two branches, and metallicity runs along the third. On the synthetic grid the coordinates were the labels up to a smooth transformation, and the question for the sample is how much of that holds. Coordinates built from distances are defined only up to a rotation, so no axis is privileged. The test is how closely a rotation and rescaling of the coordinates can follow the labels. That is a linear map $\mathbf{c}\to A\mathbf{c}+\mathbf{b}$ from the coordinates $\mathbf{c}$ to the labels, with $A$ a matrix, which rotates and rescales, and $\mathbf{b}$ an offset. That map is the linear probe \citep{alainbengio2016}, and how much of the variance of a label it accounts for measures how linearly the coordinates follow that label.

The same measure places Armillary beside other representations of the same spectra. Figure~\ref{fig:competition} puts the one-dimensional ordering of the Sequencer \citep{baron2021}, principal component analysis \citep[PCA,][]{pearson1901,hotelling1933}, t-SNE \citep{vandermaaten2008}, UMAP \citep{mcinnes2018} and an autoencoder \citep{hinton2006}, each built on the same $202\,932$ stars, under the same linear map to $(\Teff,\logg)$. Beside them is the Kiel diagram of the same stars drawn from ASPCAP's own labels. The map is fitted by least squares on all the stars and scored by $R^2$, separately for $\Teff$ and $\logg$, which is $1$ for an exact linear relation and $0$ for none.

Under the linear map Armillary's coordinates give the Kiel diagram of ASPCAP, and explain $91$ and $96$ per cent of the variance of the two axes. The autoencoder explains $76$ and $91$, PCA $75$ and $91$, UMAP $64$ and $88$ and t-SNE $69$ and $76$. The Sequencer ordering, one number per star, explains $15$ and $56$, and its map is a line through the Kiel diagram, since one number can hold one dimension of the variation. A linear map of Armillary's coordinates therefore recovers more of the sample's temperature and gravity variation than any of the other representations.

The Sequencer ordering is built on Armillary's own graph. It ranks stars by their distance along the graph's minimum spanning tree from one end of its longest chain. PCA, t-SNE and UMAP are built on the same normalised spectra $f_i$ that enter Armillary's distance, with masked pixels set to the pixel's mean over the sample, under Euclidean distances. PCA is kept to four components, the dimension Armillary uses, fitted on $30\,000$ of the stars and applied to all. UMAP is run in four dimensions, and t-SNE in three, the most its fast implementation allows at this sample size \citep{vandermaaten2014}.

The hyperparameters of t-SNE and UMAP are chosen on a random draw of ten thousand stars. There each is run over a grid of them, t-SNE over perplexities from $5$ to $100$ with both of its initialisations, and UMAP over $5$ to $100$ neighbours and minimum distances from $0$ to $0.5$. The values with the highest linear-probe $R^2$ are kept for each, a choice made with the labels.

The autoencoder is a multilayer perceptron trained to reconstruct each spectrum through a bottleneck of four numbers, scoring the reconstruction only at unmasked pixels. Its encoder has two hidden layers, mirrored in the decoder. Its hyperparameters, the width and the regularisation, are chosen on the same draw. The $36$ configurations are three widths, three weight decays, with and without dropout, and with and without noise added to the input at each pixel's own error. The choice is by the reconstruction error on a tenth of the spectra held out of the training, so no label enters it. The chosen configuration, $1024$ and $256$ hidden units with a weight decay of $10^{-5}$ and neither dropout nor input noise, is then trained on the whole sample. It uses the pixels that are unmasked in more than $95$ per cent of the spectra.

A quadratic map in place of the linear one can correct a smooth curvature of the coordinates but cannot undo a fold or a break in them. It separates the representations of Figure~\ref{fig:competition} in which the Kiel diagram is only curved from those in which parts of it are displaced. Fitted and evaluated the same way, it raises Armillary to $96$ and $98$ per cent, the autoencoder to $90$ and $93$, and PCA to $91$ and $97$, its broadened branches being curvature. The t-SNE and UMAP maps, which preserve each star's neighbours more closely than the arrangement of the neighbourhoods, reach $84$ and $82$, and $92$ and $96$. A quadratic map of Armillary's coordinates still recovers more of the temperature and gravity variation than any of the other representations.

\subsection{Transferring the labels from a few labelled stars}
\label{sec:showcase}

Beyond placing the stellar parameters without a label, a representation is useful if a few labelled stars suffice to give every other star its labels, the few-shot setting of deep learning. Here the labels of $n$ stars, the training set, are transferred through Armillary's graph, and the result is evaluated against ASPCAP on every other star. Throughout, the error of a label is a $1\sigma$ error, half the width of the central $68$ per cent of its differences from ASPCAP. It is the width of a Gaussian of the same core, which the few stars the graph places wrongly do not inflate.

A training set drawn at random follows the sample's own density, so the metal-poor giants and the high-$\alpha$ sequence, which are few, are the last regions it reaches. Until it does, the label transfer has no training star among the metal-poor giants and places them off the giant branch, on its hot side.

The coordinates exist before any label, so in practice they can decide which stars to follow up to obtain labels. The training set is drawn with probability proportional to the local density of the sample over the coordinates to the power $-1/4$, the density taken from the distance to the twentieth nearest star. This moves training stars toward the sparse regions. This matters most for the smallest training sets, in the extreme case of $32$ stars. The exponent was set by a sweep on both surveys, and larger exponents place the training stars at the edges of the sample and reduce the precision at every training-set size.

Figure~\ref{fig:showcase} shows the sample with labels transferred from training sets of $32$, $128$, $512$ and $2048$ stars, as the Kiel diagram and as the $\alphafe$--$\feh$ plane of the giants ($\logg<3.5$). The two sequences of the Galactic disc are seen in the solar neighbourhood \citep{bensby2003,adibekyan2012,anders2018} and across the Galaxy \citep{nidever2014,hayden2015}. From every training set the Kiel diagram shows the main sequence, the subgiant branch and the giant branch with its metallicity gradient. The two disc sequences are distinct from $128$ stars, and from $32$ they merge into a single tilted locus.

The training stars, marked by star symbols on the panels, explain the merging. At $32$ every training giant below a metallicity of $-0.45$ lies on the high-$\alpha$ sequence, so the low-$\alpha$ stars there take their $\alphafe$ from high-$\alpha$ training stars and the band tilts. At $128$ the training giants cover both sequences down to a metallicity of $-0.6$ and the two separate. The label transfer takes labels to vary smoothly along the graph. The two sequences are two sheets of the manifold with few stars between them and at their metal-poor ends. Until a training star lies on each sheet in a region, the label transfer interpolates across the gap there as it would across any sparse region.

The Kiel diagram shows the same sparse coverage. At $32$ and $128$ training stars the metal-poor giants split off from the giant branch as a separate arm on its hot side, where ASPCAP's diagram is continuous. Only $2$ of the $32$ training stars and $5$ of the $128$ are giants below $\feh=-1$, so the labels of the whole strand rest on a few stars. By $512$ training stars the arm rejoins the branch. The hot end of the main sequence is short at $32$ and $128$ training stars for the same reason. The stars that ASPCAP places above $6000$\,K receive $5850$\,K at the median from $32$ training stars. A star at the edge of the sample has its neighbours, and so its training stars, on one side only.

Figure~\ref{fig:onetoone_apogee} shows each transferred label against ASPCAP's on every star not in the training set. Against ASPCAP, from a training set of $32$ stars the $1\sigma$ errors are $100$\,K, $0.18$, $0.083$ and $0.049$\,dex. From $128$ they are $67$\,K, $0.10$, $0.053$ and $0.038$\,dex. From $512$ they are $44$\,K, $0.07$, $0.035$ and $0.028$\,dex. From $2048$ they are $33$\,K, $0.06$, $0.026$ and $0.022$\,dex. The illustrated draw was selected for its recovered diagrams. Across three fixed draw seeds, the $32$-star temperature error ranges from $100$ to $147$\,K, narrowing to $33.0$--$33.8$\,K at $2048$ stars.

The $\alpha$ enhancement lags the other labels because it changes the lines of a few elements, where the stellar parameters change the whole spectrum. Because the distance compares the absorption of a chunk as a whole, it separates two stars that differ in $\alphafe$ alone less readily than in the stellar parameters. The label transfer needs more training stars to resolve that difference.

Figure~\ref{fig:ladders} compares Armillary with the other representations on all four labels, over training sets from $32$ to $8192$ stars, a factor of four apart. The same training set is given to each representation. Armillary transfers its labels through the graph. Each other representation is read by a multilayer perceptron from its output to the labels, with two hidden layers of $128$ units, fitted on the training set to convergence. Its weight decay is swept from $10^{-4}$ to $10^{-1}$, and the value of lowest error on the evaluated stars is kept, separately for each of the four labels.

From a training set of two thousand stars Armillary is the most precise representation, and its lead over the autoencoder trained on the sample grows with the training set. PCA, t-SNE and UMAP are less precise than both, and the Sequencer ordering is the least precise. Armillary therefore matches the neural network from a few dozen training stars and leads it from two thousand. The autoencoder stores $13$ million weights, $53$\,MB, and trains in $8$ minutes on one NVIDIA H100 GPU. Armillary stores a sparse weight matrix of $240$\,MB, built in $45$ minutes on $24$ CPU threads (Section~\ref{sec:scaling_method}).

Every error above is measured against ASPCAP, whose labels carry errors of their own, and so includes them. How repeatable Armillary's labels are on their own is measured on the $11\,252$ stars observed twice. Their second spectra were reduced independently and are added to the sample for this test. Each of these stars then enters twice and must receive the same coordinates and the same labels from either spectrum. The median separation of the two coordinate vectors, divided by $\sqrt{2}$ and by the median distance to the tenth nearest neighbour, falls from $2.99$ to $1.81$ after refinement.

For the labels, a training set of $1024$ stars is drawn as above from the stars observed once, so that neither spectrum of a twice-observed star is in it. Every twice-observed star then receives two transferred labels, one from each spectrum. Dividing the $1\sigma$ scatter of the pair differences by $\sqrt{2}$ gives an internal repeatability of $27$\,K, $0.04$, $0.018$ and $0.011$\,dex. The same training set measured against ASPCAP gives $38$\,K, $0.07$, $0.029$ and $0.023$\,dex. The corresponding pair scatter divided by $\sqrt{2}$ for ASPCAP's own two determinations, one from each spectrum, is $18$\,K, $0.03$, $0.012$ and $0.009$\,dex. The two transferred labels share a graph and training set, so their errors can be correlated. This repeatability does not measure independent absolute accuracy.

\section{Application to DESI Spectra}
\label{sec:surveys}

The DESI Milky Way Survey \citep{cooper2023} observes stars at a resolving power of $2000$ to $5000$ over $360$--$980$\,nm, in three cameras called b, r and z \citep{desi2022}. Stellar parameters, and abundances above all, are harder to extract at this resolution, although the information is there \citep{ting2017lowres}. Data-driven models recover individual abundances from LAMOST and DESI spectra \citep{ting2017lamost,xiang2019,zhang2024}. They train on labels from a high-resolution survey, for DESI mostly through LAMOST labels tied to APOGEE. A prior from model spectra keeps the learned effect of each element on the pixels where model spectra say it has lines. A model trained on the labels alone can otherwise read an abundance from whatever correlates with it in the training set.

Nothing in Armillary is specific to APOGEE's H band. The question is whether a label-free map of DESI spectra, with APOGEE labels transferred through it, gives the Kiel diagram and the two $\alpha$ sequences with no model of the spectrum. We also ask from how many training stars.

From Data Release 1 \citep{desidr1} we take the $4799$ stars of the APOGEE sample that DESI observed with a successful, unflagged fit in the survey's stellar pipeline \citep{koposov2024,koposov2026} and a signal-to-noise above $100$. This selection uses the catalogue's red-camera signal-to-noise, $\mathrm{SN_R}$. To them we add $100\,000$ DESI stars without an ASPCAP label. These stars are above the same signal-to-noise, and by the survey pipeline's own $\Teff$, $\logg$ and $[\mathrm{Fe/H}]$ they lie within the region that the cross-matched stars cover. The graph then has many stars where the label transfer is evaluated. Only the $4799$ carry ASPCAP labels. Their metallicity label is ASPCAP's $[\mathrm{Fe/H}]$, which differs from its $\feh$ by $0.007$\,dex ($1\sigma$) for these stars.

The spectra are shifted to rest frame with the survey's own velocity and resampled uniformly in $\ln\lambda$. On that grid a pixel spans the same velocity interval at every wavelength. Each camera's flux and native inverse-variance profile are interpolated linearly, with the latter supplying the weights when cameras overlap. DESI's spectra come in calibrated flux \citep{guy2023}, so the normalisation that APOGEE's pipeline supplied is done here by a running continuum, the $90$th percentile of the flux over a window of $100$ pixels in each camera.

The interpolated inverse-variance profile, combined across cameras, also supplies the relative distance weights. We mask the telluric bands and then fit the local continuum as for APOGEE, over windows of $20$ to $60$\,nm. We build the chunks per camera in four chunkings of $10$, $20$, $40$ and $80$ chunks, of equal width in $\ln\lambda$, so that a chunk spans a similar velocity interval wherever it lies. A chunk is $47$\,nm wide at $500$\,nm in the coarsest chunking and $6$\,nm in the finest.

The same chain places the $104\,799$ DESI stars with the settings of the APOGEE run. The sweep of the neighbourhood and of $\mu_{\rm y}$ repeated on DESI gives APOGEE's setting an error within $6$ per cent of the best setting's, so it is kept. Only the $4799$ cross-matched stars can be evaluated, so the label transfer is measured on them. A training set is drawn as on APOGEE, with probability proportional to the local density of the sample over the coordinates to the power $-1/4$. The rest receive their labels through Armillary's graph of all $104\,799$ stars.

The cross-matched stars are spread more evenly over the coordinates than the APOGEE sample is over its own, so the draw changes less here. The draw still places a training star near the metal-poor dwarfs of the halo, which are few among the cross-matched stars and many in the survey. A random training set of fewer than a thousand stars leaves them to be placed off the main sequence.

Most of the survey is fainter, with a median signal-to-noise near $23$ in catalogue $\mathrm{SN_R}$, and the distant halo stars of most interest are fainter still. Fainter DESI stars have no cross-match to supply labels, so the sample is degraded instead. Independent Gaussian noise, scaled to each pixel's error, is added to lower the red-camera signal-to-noise to nominal levels of $100$, $80$, $60$, $40$ and $20$. The graph is rebuilt and the labels transferred at each level, with the same spectra and training stars throughout.

\begin{figure*}[t]
\centering
\includegraphics[width=\textwidth]{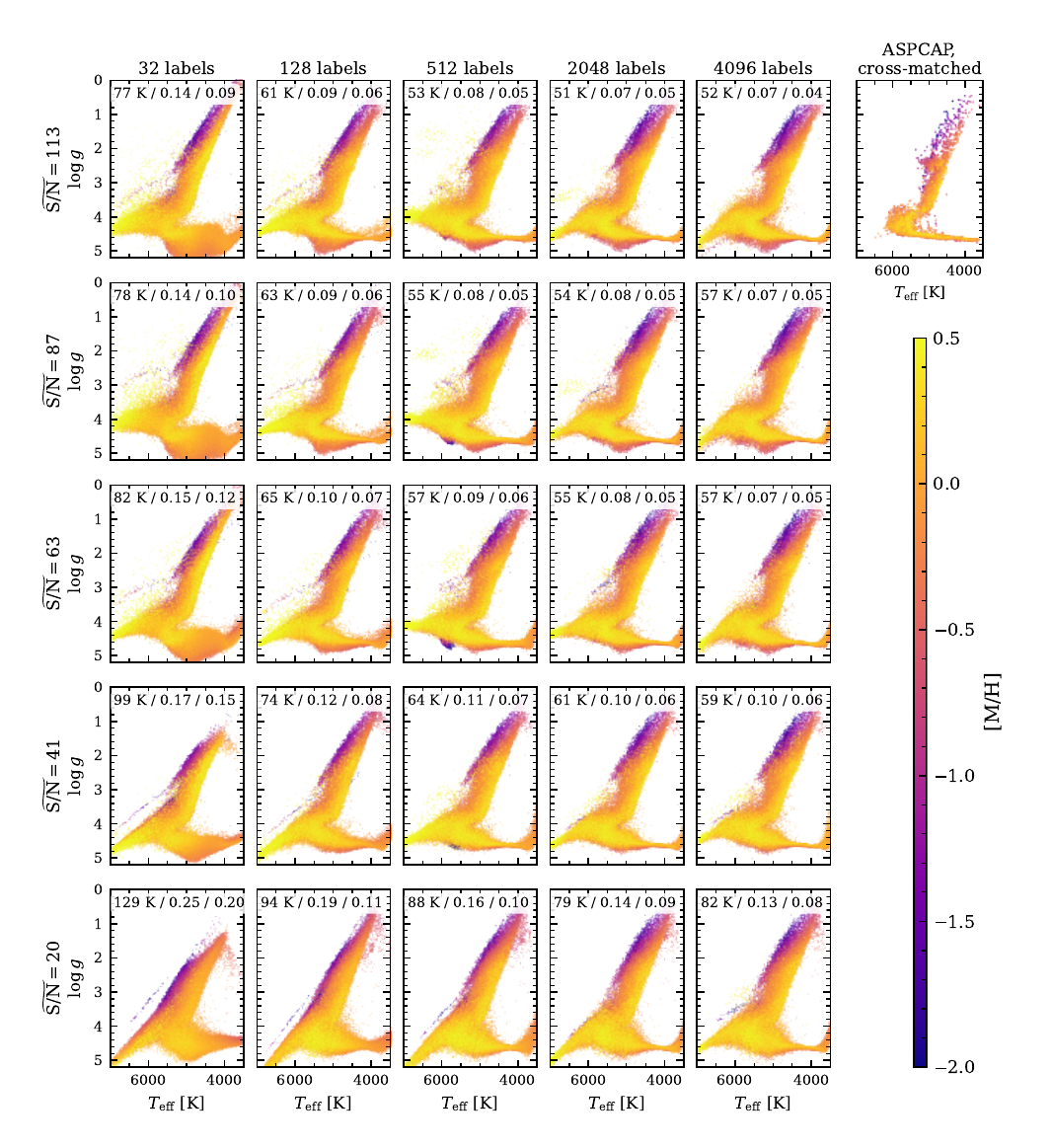}
\caption{APOGEE's labels transferred to the DESI spectra through Armillary's graph, against the median red-camera signal-to-noise (rows) and the size of the training set (columns). Each panel is the Kiel diagram of the $104\,799$ DESI stars, coloured by $\feh$, with the labels transferred from a training set of the cross-matched stars. The panel above the colour bar is the Kiel diagram of the $4799$ cross-matched stars in ASPCAP's labels. The numbers are the $1\sigma$ errors against ASPCAP in $\Teff$, $\logg$ and $\feh$ over the cross-matched stars not in the training set.}
\label{fig:surveys}
\end{figure*}

\begin{figure*}[t]
\centering
\includegraphics[width=\textwidth]{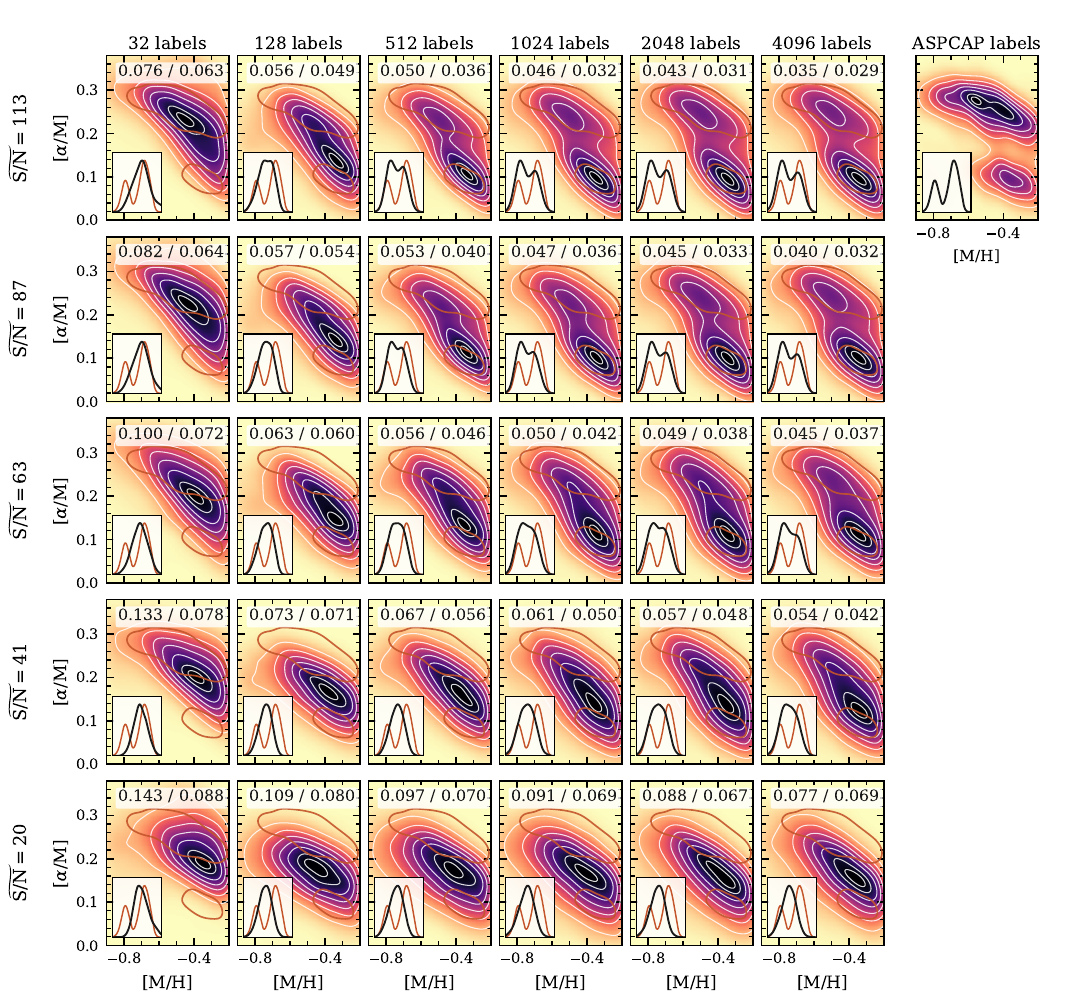}
\caption{The emergence of $\alpha$ structure in DESI against the signal-to-noise of the spectra and the size of the training set, in the $\alphafe$--$\feh$ plane of every giant of the graph by its transferred labels. Down the figure the noise increases to the median signal-to-noise named at the left, and the labels are transferred again through Armillary's graph. Across it the number of training stars increases through $32$, $128$, $512$, $1024$, $2048$ and $4096$. The panel at the top right is ASPCAP's labels for the cross-matched giants. Each panel is a smoothed density with white contours at the same fractions of its peak throughout, and the orange contour repeats one contour of the ASPCAP panel. The inset of each panel is the distribution of $\alphafe$ of the giants with $-0.55<\feh<-0.3$, ASPCAP's in orange. The numbers at the top are the $1\sigma$ errors of the transferred $\feh$ and $\alphafe$ against ASPCAP's over the cross-matched giants not in the training set.}
\label{fig:alphaseq}
\end{figure*}

Here signal-to-noise is the median flux divided by its error over usable red-camera pixels, after the noise is added. Its median over the stars is $148$ as observed, and $113$, $87$, $63$, $41$ and $20$ at the five degraded levels. Figures~\ref{fig:surveys} and \ref{fig:alphaseq} label each level by this median.

\begin{figure*}[t]
\centering
\includegraphics[width=\textwidth]{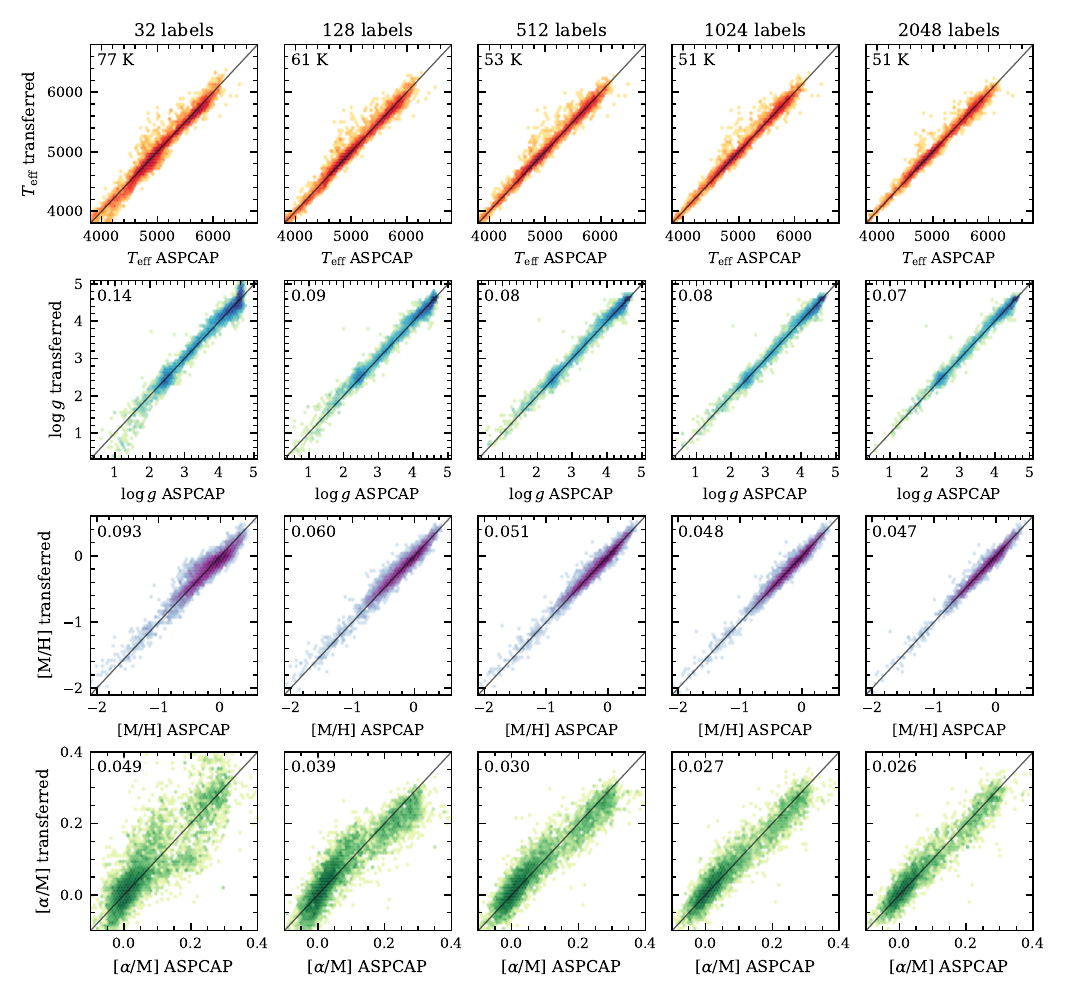}
\caption{The labels transferred to the cross-matched DESI stars against ASPCAP's at signal-to-noise about $100$. Each row is one label, and each column is one training set of $32$, $128$, $512$, $1024$ and $2048$ stars. The panels show the cross-matched stars not in the training set. Each panel is the density of the stars on a logarithmic scale, with the one-to-one line, and its number is the $1\sigma$ error of the difference. All four rows use the same held-out stars, including dwarfs and giants.}
\label{fig:onetoone}
\end{figure*}

Figure~\ref{fig:onetoone} shows each transferred label against ASPCAP's at signal-to-noise about $100$, training set by training set. Its panels contain roughly fifty times fewer held-out stars than Figure~\ref{fig:onetoone_apogee}, so sparse tails are less visible, and the quoted errors provide the quantitative comparison. $\Teff$, $\logg$ and $\feh$ follow the one-to-one line to $1\sigma$ errors of $77$\,K, $0.14$ and $0.093$\,dex from $32$ training stars and $53$\,K, $0.08$ and $0.051$\,dex from $512$. The three stellar parameters, at least, are transferred from a few dozen training stars. From merely $8$ training stars the errors are $117$\,K, $0.21$ and $0.21$\,dex, the Kiel diagram coarse but recognisable, although not shown.

Figure~\ref{fig:surveys} shows the Kiel diagram of the whole graph at each signal-to-noise and training set. The giant branch is in place from the smallest training set at every signal-to-noise tested. A broad band at first, the main sequence narrows with the size of the training set and broadens with the noise. The errors from $1024$ training stars rise with the noise from $51$\,K, $0.08$ and $0.048$\,dex at signal-to-noise about $100$ to $84$\,K, $0.15$ and $0.089$\,dex at about $20$. From $32$ training stars at signal-to-noise about $20$ the errors are $129$\,K, $0.25$ and $0.203$\,dex and a streak of stars runs from the turn-off to the giant branch. The recovery of the Kiel diagram from small training sets down to a signal-to-noise of about $20$ indicates that the graph already captures much of the stellar-parameter structure before the labels are attached.

For comparison, the survey's own pipeline labels, as published and without the metallicity calibration that \citet{koposov2026} provide, differ from ASPCAP. On the same held-out stars used for the $1024$-star label transfer, the differences are $108$\,K, $0.26$, $0.15$ and $0.06$\,dex. Here we compare the pipeline's $[\mathrm{Fe/H}]$ and $[\alpha/{\rm Fe}]$ to ASPCAP's $[\mathrm{Fe/H}]$ and $\alphafe$. The label transfer from $1024$ training stars at signal-to-noise about $100$ differs by $51$\,K, $0.08$, $0.048$ and $0.027$\,dex. The transferred labels lie closer to ASPCAP than the survey's own pipeline does, as expected of a label transfer that takes ASPCAP's labels as its training set. The two surveys then share one label scale.

$\alphafe$ needs a larger training set. As on APOGEE, the $\alpha$ enhancement changes the lines of a few elements where the stellar parameters change the whole spectrum. Here at low resolution those lines are blended with the rest, so the label transfer needs the most training stars for this label.

Figure~\ref{fig:alphaseq} shows the $\alphafe$--$\feh$ plane of every giant of the graph by its transferred labels, at each signal-to-noise and training set. Each panel has an inset with the distribution of $\alphafe$ of the giants with $-0.55<\feh<-0.3$, where both of ASPCAP's sequences are present. From the smallest training sets the range of $\alphafe$ is compressed, the high-$\alpha$ stars pulled down and the low-$\alpha$ stars up, and the compression eases with the size of the training set. At signal-to-noise about $100$, two branches emerge from about $512$ training stars and become more apparent with larger sets.

The noise weakens the branches and raises the errors. At $1024$ training stars the errors of $\feh$ and $\alphafe$ in giants rise from $0.046$ and $0.032$\,dex at about $100$ to $0.091$ and $0.069$\,dex at about $20$. At about $80$ the branches emerge from the same training set size of $512$ but are shallower. At about $60$ there are hints at $512$ and definite shoulders from $2048$. At about $40$ a tentative hint of bimodality remains at $4096$ training stars. The label-free map of DESI therefore gives the stellar parameters from a few dozen training stars and the two $\alpha$ sequences from about five hundred, at signal-to-noise above about $60$.

\section{Discussion}
\label{sec:discussion}

\subsection{Deep representation learning}
\label{sec:representation}

Before any label is attached, a set of spectra has a structure of its own, since a few parameters generate the spectra and they lie on a manifold of a few dimensions. A representation that recovers that manifold from all the spectra lets a few labelled stars supply the labels of every other one. Deep learning makes that aim practical. Self-supervised neural networks now learn from unlabelled spectra first and take labels afterwards \citep{sedaghat2021,specclip2026}, and foundation models trained on spectra and labels together are adapted to new tasks \citep{leungbovy2024,spectrafm2024}. A deep neural network can learn flexible nonlinear relations. It returns, however, a latent representation in which the physical parameters are often entangled, and which usually needs labels for a physical interpretation.

The question asked here is how much of that structure classical machine learning tools recover when they are given the right ingredients. Armillary defines a distance between spectra that respects the physics of absorption, builds the graph of nearest neighbours under it, takes coordinates from the shortest paths through the graph, and transfers labels through it. On APOGEE the four coordinates follow the three stellar parameters, temperature, gravity and metallicity, closely under a linear probe. On both surveys the graph transfers the stellar parameters from a small training set.

\subsection{Classical manifold methods}
\label{sec:uses}
Armillary follows the key ideas of classical manifold learning. The coordinate step is Isomap \citep{tenenbaum2000}, the refinement's weights are those of locally linear embedding \citep{roweis2000}, and the label transfer is the label propagation of semi-supervised learning \citep{zhu2002,zhou2004} in its form through locally linear weights \citep{wang2008}. The Sequencer \citep{baron2021} applied the Wasserstein distance of optimal transport, among other distances, to spectra and showed that a single ordering built on them follows temperature. Armillary extends that distance to signed absorption and retains the graph-based construction. It replaces the ordering, which follows a single dimension, by coordinates on every dimension along which the spectra vary, and builds the refinement and the label transfer on the same graph.

Isomap, locally linear embedding and label propagation predate deep learning. On stellar spectra, authors use label propagation to find rare objects \citep{si2014}. They use Isomap and locally linear embedding, and later t-SNE and UMAP, for classification and morphology \citep{daniel2011,matijevic2012,bu2013,bu2014,traven2017,matijevic2017,kao2024}. PCA's projection \citep{bailerjones1998,mcgurk2010} and the Sequencer's ordering have likewise been used before any label enters, and Armillary's coordinates can be used the same way because it is unsupervised. Stellar spectra vary mainly with the stellar parameters and the $\alpha$ enhancement, so the stars are where the construction can be tested against existing labels.

In Armillary the physics enters through the distance. The Wasserstein distance compares where the absorption lies along the wavelength axis once the total amount of absorption is divided out. A step in temperature or gravity gives a similar distance for a cool star crowded with lines and for a hot star with few. A small shift in wavelength changes the distance only slightly. The Euclidean distance is dominated by how much absorption a spectrum contains, so the same step is large for the line-rich star and small for the line-poor one. A graph built on it is stretched in one corner of the label space and compressed in another, and the coordinates inherit that unevenness. The Wasserstein distance makes the geometry of the sampled manifold more even in these tests, improving the local distances on which the graph and coordinates depend.

Beyond the samples analysed here, the coordinates of quasar or galaxy spectra, shaped by more than one process, can show along which dimensions a sample varies at all. The same holds for white dwarf spectra or stellar light curves, each with a distance that respects its own physics. Authors apply manifold methods to the spectra of galaxies, quasars and white dwarfs \citep{vanderplas2009,richards2009,kaopasj2024,kao2024}. A spectrum far from every neighbour, from a reduction defect or a peculiar object, shows as such on the same manifold \citep{matijevic2012}.

\subsection{Combining Armillary with spectral model fitting}
\label{sec:synergy}
The supervised data-driven models complement Armillary. The Cannon \citep{ness2015,casey2016} and DD-Payne \citep{ting2017lamost,xiang2019} learn a function from labels to spectra. Trained on thousands of labelled stars, as in their survey-scale applications \citep{casey2016,xiang2019}, they fit the whole spectrum of each star and recover the abundances of individual elements. Armillary needs a few dozen to a few hundred labelled stars. Where labelled stars are few, as for a new instrument or a wavelength range with few reference stars, the graph transfers their labels from the handful that exist. The individual elements are out of its reach as it stands, because of the distance. It compares the shape of the absorption over chunks of a few nanometres. It therefore resolves what many lines carry together, the stellar parameters and the $\alpha$ enhancement, before an element with one or two weak lines.

The $\alpha$ enhancement already shows where that limit lies. The label transfer of the stellar parameters needs a few dozen training stars. That of the $\alpha$ enhancement needs a hundred on APOGEE and five hundred on DESI at signal-to-noise above about $60$. The difference reflects the weaker imprint of the $\alpha$ lines on the distance, which weighs them with every other line of the chunk. Weighting the pixels that a line list singles out can improve sensitivity to one element. A distance made sensitive to every element would, however, reduce to fitting a physical model of the spectra. That would defeat the aim of this study, which is to find how far a purely data-driven, unsupervised construction can go.

\subsection{Limitations}
\label{sec:limitations}

Armillary's limits follow from its construction. The graph is built once, from the distances among all the spectra of a sample, and the labels are transferred through it. A new spectrum therefore has no place in it until the construction is rerun, and there is no fitting of one spectrum at a time. For a sample analysed as a whole this matters little. For a stream of new spectra one route is the landmark placement of the implementation, which positions a star from its distances to a few hundred landmarks without rebuilding the graph \citep{desilva2004}. Its labels would then come from the neighbours it is placed among.

Sample size is the other side of the same limit, since the neighbour search dominates the computational cost and grows with the length of the feature vectors as well as with the number of spectra. A graph of several million spectra would therefore need shorter features, or a search on a subset with the rest placed through the landmarks. The graph also describes only the sample it was built from, a limit it shares with other data-driven models. A star of a kind the sample does not otherwise contain, a binary, a chemically peculiar star or a spectrum with a reduction defect, is placed wherever its spectrum is closest. It receives the labels of that neighbourhood, and only the geometry on the manifold warns of it.

The noise is handled less completely than in a fit. A model fitted to a spectrum, as the Cannon or the Payne is, weighs each pixel by its own error. The cumulative curve averages the pixel noise within a chunk, and the reliability profile discounts the noisiest pixels, but neither weighs one pixel against another as a fit does.

Mixing stars of widely different signal-to-noise can also distort the neighbourhoods used for label transfer. Noise perturbs the distances, and wavelength-dependent noise can introduce structure that differs from star to star. The experiments therefore apply one noise prescription to the same sample at each level, although the achieved signal-to-noise still varies across stars. The label transfer of the stellar parameters and the $\alpha$ enhancement degrades gradually with the signal-to-noise, as Figures~\ref{fig:surveys} and \ref{fig:alphaseq} show.

\section{Conclusion}
\label{sec:conclusion}

Most representation learning on spectra now uses deep neural networks, which learn a latent space from the unlabelled spectra and capture flexible nonlinear relations, although interpreting their latent representation usually requires labels. We ask, alongside that approach, how much of the structure of a stellar spectral sample the classical tools of manifold learning recover when they are given the right ingredients. The construction that answers it and its main results are summarised below.

\begin{itemize}
\item The distance decides the geometry of the manifold of spectra. The Wasserstein distance between normalised absorption profiles compares where absorption lies once the total amount of absorption is divided out. It makes the manifold more even than the Euclidean distance does. The Euclidean distance stretches one corner of the manifold and compresses another.
\item Under that distance the Sequencer's ordering generalises to the whole manifold. Shortest paths through the neighbour graph give coordinates in as many dimensions as the eigenvalues single out, in place of the one dimension an ordering can follow. A label-free refinement and the transfer of labels are built on the same graph.
\item The coordinates come closest to temperature and gravity under a linear map. A linear map of the coordinates gives the Kiel diagram, recovering $91$ per cent of the variance of $\Teff$ and $96$ per cent of that of $\logg$. This is more than PCA, t-SNE, UMAP or a tuned autoencoder give under the same map, so the geometry of the manifold carries the parameters before any of them is fitted.
\item Given the same training sets, Armillary's label transfer is more precise than PCA, t-SNE, UMAP and the Sequencer ordering, and comparable to a tuned autoencoder. On the stellar parameters and the $\alpha$ enhancement the classical chain rivals representation learning with a neural network, and it is interpretable.
\item A few dozen training stars recover the stellar parameters on APOGEE and on DESI. From $32$ training stars the Kiel diagram of either survey is recognisable. The $1\sigma$ errors against ASPCAP are $100$\,K, $0.18$ and $0.083$\,dex on APOGEE, and $77$\,K, $0.14$ and $0.093$\,dex on DESI at signal-to-noise about $100$. They improve to $33$\,K, $0.06$ and $0.026$\,dex from $2048$ stars on APOGEE and $51$\,K, $0.07$ and $0.047$\,dex from $2048$ on DESI. The two $\alpha$ sequences of the disc are distinct in giants from a hundred training stars on APOGEE, and on DESI they emerge from about five hundred at signal-to-noise above about $60$.
\end{itemize}

Because the coordinates are built from the spectra alone, Armillary can be applied to the archive of any spectrograph before a single label exists. The coordinates show the dimensions along which a sample varies and the objects that lie apart from the rest. A few labelled stars, for example those shared with another survey, are then enough to place every other star on that survey's label scale, and so to homogenise the labels across surveys.

Armillary can suit the broad, low-resolution spectra of Gaia \citep{gaia2023}, Euclid \citep{euclid2025}, the Roman Space Telescope \citep{roman2019} and SPHEREx \citep{spherex2026}, which are arriving, or about to arrive, in vast numbers. Its distance compares the broad shape of the absorption, which is what such spectra carry. Fast, label-free classical methods of this kind can be a practical way to extract stellar parameters at that scale.

Beyond these applications, a distance that respects the physics gives a manifold that classical machine learning tools recover, so for tasks of this kind a black box can be unnecessary. Neural networks, foundation models among them, can be trained under such a distance or held to the geometry it defines, and so return a latent representation that can be read. Understanding what such simple ingredients recover is a step toward understanding what large deep learning models learn from spectra.

The code and the data products of this paper are archived on Zenodo \citep{tingsaad2026zenodo}.

\begin{acknowledgments}
We thank Todd A. Thompson, Christopher S. Kochanek, Krzysztof Z. Stanek and Xiaosheng Zhao for useful discussions, and Tomasz R\'o\.za\'nski and Jiadong Li for their work with YST on the earlier attempts described below. YST acknowledges a Humboldt Research Fellowship from the Alexander von Humboldt Foundation, held at the Max Planck Institute for Astronomy.

This project is the accumulation of several years of YST's failed attempts to make the Sequencer and autoencoders say something useful about stellar spectra, for the reasons laid out in this paper. It came back to life during work in preparation with Xiaosheng Zhao on foundation models built on joint-embedding predictive architectures. Their success at learning cleaner representations suggested that a clean representation of stellar spectra was possible, and left the question of whether it could also be understood. The way in was inspired by a discussion with Todd A. Thompson, Christopher S. Kochanek and Krzysztof Z. Stanek on distances in the spirit of the Wasserstein distance and on which spectra should count as adjacent. The manifold ideas are older. Isomap and locally linear embedding date back to before graduate school, when YST was trying to measure the dimensionality of chemical abundance space with principal component analysis \citep{ting2012}. Some of them were explored then, and came back more than a decade later.

The code and much of the brainstorming were done with GPT-5.6, GPT-6, Claude Opus 5, Claude Opus 5.1 and Claude Opus 5.5. The paper began as an extensive outline by the authors. The authors wrote the first draft of the introduction, motivation, discussion and conclusion. Claude drafted the method and the results, where holding every ablation detail made it more accurate, and did the copy editing throughout, although left to itself it writes Claudish rather than English, verbose and rarely the more helpful for it. The language models also canvassed the literature, which helped to map what had already been done. The direction of the project, and every result and sentence of the paper, were set or validated by the authors.
\end{acknowledgments}

\software{Armillary \citep{tingsaad2026zenodo}, NumPy \citep{harris2020}, SciPy \citep{virtanen2020}, scikit-learn \citep{pedregosa2011}, Numba \citep{lam2015}, PyNNDescent, umap-learn \citep{mcinnes2018}, openTSNE \citep{policar2024}, PyTorch \citep{paszke2019}, Astropy \citep{astropy2013,astropy2018,astropy2022}, Matplotlib \citep{hunter2007}}

\bibliographystyle{aasjournal}
\bibliography{references}

\end{document}